\documentclass[eqsecnum,aps,citesort]{revtex4}

\usepackage{graphicx}
\usepackage{amssymb}

\newcommand{\be}{\begin{equation}}
\newcommand{\ee}{\end{equation}}
\newcommand{\ba}{\begin{eqnarray}}
\newcommand{\ea}{\end{eqnarray}}
\newcommand{\ban}{\begin{eqnarray*}}
\newcommand{\ean}{\end{eqnarray*}}
\newcommand{\p}{\partial}
\newcommand{\f}{\frac}

\begin{document}


\title{A Continuum Description of Rarefied Gas Dynamics (III)---
The Structures of Shock Waves}
\author{Xinzhong Chen}
\address{Astronomy Department, Columbia University, New York, NY 10027}
\author{Hongling Rao}
\address{Microeletronics Sciences Laboratories, Columbia University,
New York, NY 10027}
\author{Edward A. Spiegel}
\address{Astronomy Department, Columbia University, New York, NY 10027 }

\date{\today}

\begin{abstract}

\noindent We use the one-dimensional steady version of the equations
derived in paper I to compute the structure of shock waves.  The agreement
with experiment is good, especially when we retain the experimental
value of the Prandtl number adopted in II.

\end{abstract}


\maketitle

\section{Introduction}

In the simplest descriptions of shock waves, one uses the Euler
equations of fluid dynamics to develop jump conditions across the
discontinuity that describes the shock.  These conditions have
qualitative value but there are problems for
which such limiting solutions are not adequate.  For example, in
studying radiation from shock waves, say to infer the properties
of the radiating atoms, the conditions within the shocks may
become important.  If the matter ahead of the shock is neutral and
that behind the shock is fully ionized, the observed spectral lines
may be formed in the shock itself and so conditions there need to be
carefully worked out. This example is one of many that we might have
cited to motivate our present study of the structure of shock waves. Even
a cursory look at the literature on this problem gives a clear
picture of the importance that has been attached to it.

Much of the work on the structure of shock waves has been aimed at
improving on the description provided by the Navier-Stokes
equations. A review of such attempts was given by Galkin {\em et.
al.}~\cite{gal98}, who compared various higher order solutions of
Boltzmann equation with results from the Chapman-Enskog
~\cite{cha70} and Hilbert methods ~\cite{gra63}. In terms
of shock structure alone, the higher order solutions give a
significantly improved result~\cite{zho93,fis89}, but a number of
fundamental problems concerning the status of these equations
remain open ~\cite{gra63,gal98,kli92,woo93}.  It is especially
troubling that a large number of higher order nonlinear terms in
some of the proposed improvements make it difficult to use the
results in realistic problems.  Similar difficulties beset
extensions of Grad's moment method~\cite{gra49,gra63}, such as the
Extended Irreversible Thermodynamics~\cite{mul98}, which does give
a good result for the problem of ultrasonic sound wave
propagation. It remains true however, that a large number of
equations must be solved to achieve a reasonable accuracy with
EIT.

When the densities are sufficiently low, direct numerical simulation
by the Monte Carlo method is the most reliable way to compute
high-Knudsen-number flows, though the computational cost may be high in
regimes near continuum limits~\cite{myo99}.  A relatively new and
effective method that has been described for direct solution of the
relaxation (or BGKW) model does offer hope of improvement~\cite{pre93}
but it remains to taken beyond the two-dimensional problems that it has
so far tested very well on.

It thus appears that an effective macroscopic description would be
of value for such problems and so, in this paper, we
show how the equations derived in paper I may fill this need.  To
do this, we work out the structure of shock waves in one-dimensional
steady flow starting from the equations of paper I.  Those equations
were derived from kinetic theory without using some of the traditional
simplifications associated with the Chapman-Enskog approach.   In
particular, we did not use results from lower order approximations
to simplify higher order equations.  In paper II, the first-order
development (in mean free path) of paper I was tested against
observations of ultrasound propagation.  We found good quantitative
agreement between our theory and the experiments in the high Knudsen
number regime where Navier-Stokes equations clearly fail.  Though the
success of our approach works well beyond the expected limits of validity
of the theory, such occurrences are not unheard of in good asymptotic
methods.

In the present paper, we go on to see how well the theory works for the
computation of the structure of shock waves.  Since the shock thickness
is typically of the order of a mean free path, we again are
pushing against the limits of validity of the expansions used in deriving
the fluid equations.  We confront in addition the added challenge of
strong nonlinearity.  Moerover the flow considered in shock theory can
is far from thermodynamic equilibrium, so this too makes for a stringent
test.

A simplification that makes the comparison relatively easy to draw is
that, in the study of shocks, we may separate the continuum differential
equations of fluid motion from the boundary conditions that must be
stated to complete a well-posed problem. The boundary conditions for the
shock problem is not in question since we may impose the aforementioned
jump conditions, expressed in terms of the Rankine-Hugoniot relations, to
serve in the role of the boundary conditions. Hence in this study of shock
structure, we are able to focus on effects due only to the differential
equations themselves and so examine the validity of our version of
the fluid equations.

In Sec. II, we recall the fluid dynamical equations derived in paper I.
Then, in Sec. III, the structure of one-dimensional shock waves are
computed and compared to analogous results obtained with the Navier-Stokes
equations as well as with experiments.  The paper concluded with a
brief discussion in Sec. V.

\section{Statement of the Equations}

In paper I we proposed a modification of the usual asymptotic
techniques for deriving fluid equations from kinetic theory.  Our
procedure avoids the simplification introduced by Chapman
and Enskog~\cite{cha70} in which the results of lower approximations
are introduced into the higher approximations.  When we proceed in
this way, we obtain these fluid equations~\cite{chen00,che00a,che00}:

\ba & & \p_t \rho+\nabla\cdot(\rho{\bf u})=0 \label{x13} \\ & &
\p_t\ u+{\bf u}\cdot\nabla{\bf u}+\f{1}{\rho}\nabla\cdot{\mathbb P} =0
\label{x14} \\ & & \p_t T+{\bf u}\cdot\nabla T+\f{2}{3\rho R}({\mathbb P}:
\nabla{\bf u}+\nabla\cdot{\bf Q}) =0 \ ,
\label{x15}
\ea where $\rho$ is the mass density, ${\bf u}$ is the average velocity of
the particles in a fluid element cell, and $T$ is the temperature.
We assume that the gas is made up of identical, structureless particles
with mass $m$ so that $R=k/m$ is the gas constant with $k$ the Boltzmann
constant.  Our expression for the stress tensor is \be
{\mathbb P}=\left[\rho RT-\mu{D\ln T\over Dt}
+\frac{2}{3}\nabla\cdot{\bf u})
\right]{\mathbb I}-
\mu{\mathbb E}:\nabla \nabla {\bf u}
\label{x39}
\ee where \be {D\over Dt}=\partial_t+{\bf u}\cdot\nabla \; ,
\label{d0}\ee and
\be\mu=\tau \rho RT\label{mu}\ee
is the viscosity expressed in terms of the mean free time $\tau$.  For
the stress tensor we found
\begin{equation}
E^{ij} = {\partial u^i\over \partial x_j } + {\partial u^j\over
\partial x_i } - {2\over 3} \nabla \cdot {\bf u} \, \delta^{ij} \ .
\label{E} \end{equation}
and for the heat current, we have \be {{\bf Q}}=-\eta\nabla\ln(\rho
T^{-\f{3}{2}})-\f{7}{2}\eta\nabla\ln T -\f{5}{2}\mu{D \bf u\over Dt} \ ,
\label{x40}
\ee where $\eta=\f{5}{2}\mu RT$ for the relaxation model.

By contrast, in the  Navier-Stokes equations with no bulk viscosity, one
has
\be {\mathbb{P}}=\rho RT{\mathbb I}-\mu {\mathbb E}:\nabla \nabla {\bf u}
\label{x41}
\ee and \be {{\bf Q}}=-\eta\nabla\ln T \ .
\label{x42}
\ee

\section{Shock Theory}
\subsection{Basic equations}

\begin{figure}[hbt]
\begin{center}
\includegraphics[bb=2in 4in 6in 7in,width=8cm]{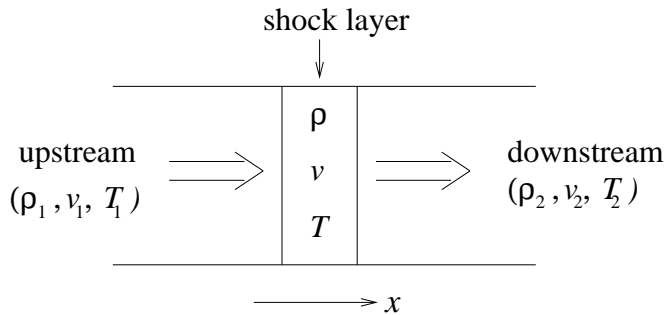}
\end{center}
\caption{The schematic diagram of a shock wave} \label{fig0}
\end{figure}

As inidcated in Fig.~\ref{fig0}, when the velocity of the flow in the
upstream ($x\rightarrow -\infty$) exceeds the sound speed of the
medium, a shock front forms.  In a frame comoving with the shock
front, we see a steady shock layer, whose structure is
determined by the upstream thermodynamic quantities and the
dissipation mechanism. The structure of the shock wave provides a
straightforward test of the equations.

Let us use subscripts $1$ and $2$ to denote values at large distances
upstream ($x\rightarrow -\infty$) and downstream ($x\rightarrow\infty$)
of the shock front, with ${\bf u} = (v(x),0,0)$, $\rho$ and $T$ as
functions of only $x$. In the frame comoving with the
shock front, we have $\partial_t=0$. On integrating (\ref{x13}) to
(\ref{x15}) from uniform upstream state to an arbitrary position $x$ in
the shock, we obtain \ba & & \rho v=\rho_1 v_1 \label{s1} \\ & &
p+\rho v^2+E^{11}=p_1+\rho_1 v_1 \label{s2} \\ & & \rho
v(c_vT+\frac{1}{2} v^2)+pv+v\pi_{xx}+q_x =\rho_1
v_1(c_vT_1+\frac{1}{2}v_1^2)+p_1 v_1 \ ,
\label{s3}
\ea where $c_v=\frac{3}{2}R$, $E_{11}$ is the $\hat x\hat x$
component of the viscous stress tensor, and $Q^{1}$ is the $\hat x$
component of the heat current.  The simple form of the right sides of
these equations results from the vanishing of the derivatives of the
fluid variables far from the shock.

From (\ref{s1}), we can express $\rho$ in terms of $v$: \be
\rho=\frac{\rho_1 v_1}{v} \label{s4} \ . \ee
Combination of (\ref{s4}) with (\ref{s2}) and (\ref{s3}) leads to \ba
E^{11} & = & R\rho_1v_1\left(\frac{T_1}{v_1}-\frac{T}{v}\right) +
\rho_1v_1(v_1-v) \label{s5} \\
q^1 & = &
\frac{3}{2}R\rho_1v_1(T_1-T)+p_1(v_1-v)+\frac{1}{2}\rho_1 v_1(v_1-v)^2
\label{s6} \ . \ea

We nondimensionlize these equations using $v_1$ as the unit of speed
$T_1$ as unit of temperature and the mean free path $\lambda_1$ at
upstream infinity as the unit of length.  Then, we introduce the
nondimensional quantities $X=x/\lambda_1$, $w(X)=v/v_1$, $\theta(X) =
T/T_1$ and \begin{equation}
\varpi(X) = {E^{11}\over p_1} \qquad {\rm and} \qquad q(X) = {Q^{1}\over
p_1 v_1}\ . \label{EQ} \end{equation}
The equations become \ba
\varpi & = &1-{\theta\over w} +\frac{5}{3} M_1^2(1-w)
\label{s7} \\
{2q\over 3} & = & 1-\theta +\frac{2}{3}(1-w)+
\frac{5}{9}M_1^2(1-w)^2
\label{s8} \ , \ea
where the Mach number is
$$M_1=\frac{v_1}{c_{1}} \qquad {\rm with} \qquad c_{1}=\sqrt{\gamma
kT_1}$$
and $\gamma=5/3$.

We see in from (\ref{mu}) that the viscosity is $\mu=\tau p$, where
$\tau$ is the mean flight time of the particles.  In turn, $\tau$ is
the mean free path over the mean speed, which is the local speed of
sound.  Here, we adopt the simplest form of the relaxation model, namely
that with constant $\tau$.  Then, we follow Gilbarg {\em et
al.}~\cite{gil53} and take
\be \mu = \frac{\lambda_1 p_1}{\sqrt{2RT_1}}s \label{s9} \ ,
\ee
where $s$ is a parameter that is adjusted according to the nature of
the constituent particles. For argon, $s=0.816$ has often been
used ~\cite{cha70}, but Kestin {\em et al.}~\cite{kes84} suggest that
$s=0.64$ is a better value for argon.  We shall adopt the latter, more
recent value in these computations.  As to the conductivity, the
relaxation model gives $\eta=\frac{5}{2}\mu RT$ but, when the Boltzmann
collision term is used, we obtain a slightly different value in better
agreement with experiment.  As the difference results from the atomic
model rather than to fluid dynamical issues, we shall adopt the formula
$\eta=15\mu RT/4$ to remove the effects
of the inaccuracy of the atomic model.

To these formulae we adjoin the the closure relations (\ref{x39}) and
(\ref{x40}) that may be rewritten in nondimensional form as \ba
\varpi & = & -\sqrt{\frac{5}{6}}M_1 s
\left[2 w'+ w \frac{\theta'}{\theta}\right] \label{s11} \\
q & = &-\frac{3}{2}\sqrt{\frac{5}{6}}\frac{s}{M_1}
\left[2\theta'+\frac{10}{9}M_1^2 w
w'-\theta\frac{w'}{w}\right] \ ,
\label{s12}
\ea where the primes indicate differentiation with respect to $X$.

On combining (\ref{s11}) and (\ref{s12}) with (\ref{s7}) and (\ref{s8}),
we obtain these equations:
\ba
-\sqrt{\frac{5}{6}}M_1 s
\left[2 w'+w \frac{\theta'}{\theta}\right] & = &
1-\frac{\theta}{w} +\frac{5}{3} M_1^2(1-w) \label{s15} \\
-\frac{3}{2}\sqrt{\frac{5}{6}}\frac{s}{M_1}
\left[2\theta'+\frac{10}{9}M_1^2 w w'
-\theta\frac{w'}{w}\right] & = & 1-\theta
+\frac{2}{3}(1-w)+\frac{5}{9}M_1^2(1-w)^2 \label{s16} \ . \ea

For comparison, we also write down the equations determining the shock
structure from the Navier-Stokes equations: \ba
-\frac{4}{3}\sqrt{\frac{5}{6}}M_1 s w' & = &
1-\frac{\theta}{w} +\frac{5}{3} M_1^2(1-w) \label{s13} \\
-\frac{3}{2}\sqrt{\frac{5}{6}}\frac{s}{M_1}\theta' & = &
1-\theta +\frac{2}{3}(1-w)+\frac{5}{9}M_1^2(1-w)^2
\label{s14} \ . \ea

\subsection{Critical Mach number}

Before carrying out the numerical integration of equations
(\ref{s15})$-$(\ref{s16}), we examine them to draw
some general conclusions.  First, we observe that, since
the $x$-derivatives of $w$ and $\theta$ vanish far upstream and
downstream, the right sides of equations (\ref{s15})-(\ref{s16})
(as well as of (\ref{s13})-(\ref{s14})) must vanish there.
These conditions provide two simultaneous equations which are readily
solved for the fixed points $(w_1,\theta_1)=(1,1)$ and
$$(w_2,\theta_2)=\left({M_1^2+3\over 4M_1^2}\,
{5M^4_1+14M_1^2-3\over 16M_1^2}\right) \, .  $$

Our aim is to find solutions that connect the two fixed points.
However, in the case of our equations, though the fixed points must
occur at the locations just found in the $(w,\theta)$ plane, those
locations need not be fixed points: the left sides of
(\ref{s15})-(\ref{s16}) may vanish when the determinant of the
matrix of the coefficients of the derivatives vanishes.  This is
seen when we write the equations in the standard form
\begin{equation}
{\mathcal M} \pmatrix{w' \cr \theta'} = -{1\over s}\sqrt{6\over 5}
\pmatrix{\varpi/M_1 \cr M_1 q} \label{deriv} \end{equation}
where \begin{equation}
{\mathcal M}=\pmatrix{2 & w/\theta \cr \xi w-\theta/w & 2} \label{matrix}
\end{equation}
with $\xi=10M^2_1/9$.

If we can solve (\ref{deriv})-(\ref{matrix}) to obtain explicit
expressions for the derivatives, we can then solve the coupled first-order
ordinary differential equations to find solutions connecting the
fixed points.  This is possible if the determinant of ${\mathcal M}$ does
not vanish. The critical condition is then obtained by setting the
determinant to zero, a step that leads to the relation \begin{equation}
\theta = {2\over 9} M^2_1 w^2\; . \end{equation}
At upstream infinity, we have $(w, \theta) = (1,1)$ and this leads to
the critical value $M_1=3/\sqrt{2}\approx 2.12$.  We cannot solve for
the derivatives when $M_1$ exceeds this value and so the theory breaks
down above this value.

When the upstream flow speed is large enough to make the Mach number
surpass the critical Mach number $M_1^c$, $D$ changes its sign. If
this leads to either $\frac{D_\omega}{D}>0$ or $\frac{D_\theta}{D}<0$
near the upstream, which is true for our case here, it will be
impossible to match the downstream values monotonically. This happens
for any hyperbolic system, as is commonly seen in moment
formalisms~\cite{mul98} and first noticed in Grad's 13-moment method.
Interestingly, it is numerically verified~\cite{mul98,wei95} that
only when the largest upstream critical Mach number is surpassed,
does the above situation happen. As more and more moments are
included, the largest critical Mach number becomes larger and larger,
which in a sense partly relaxes the constraints on the application to
shock study with moment method. It is thus expected that with the
higher-order terms included in our formalism, the range of the Mach
number will be extended.  Nevertheless, the constraints on the Mach number
is not due to the expansion scheme, but due to the nature of the
relaxational model.  In a following paper, we
shall show that, when classical Boltzmann integral is used in place of
the relaxation term, the constraints on Mach number will be
removed. However, physically, information speed can not be infinite,
which indicates that the final version of hydrodynamics must be of
hyperbolic nature. Intuitively, this issue can only be addressed
consistently under the frame of relativity. Amazingly, when our
modified Hilbert expansion is applied to relativistic Boltzmann
equation~\cite{che00}, a causal hydrodynamics which has constraints
on the Mach number originated from relativity is obtained.  Such a
hydrodynamics is free of the stability problem and causality problem
inherent in the Steward-Israel version of
hydrodynamics~\cite{ste77,isr79b}, the relativistic counterpart of N-S
formalism.  For our present purpose, we are content with the
nonrelativistic version.  The major concern here is whether the accuracy
is improved when Knudsen number is not small.

\subsection{Comparisons}

We integrate the equations~(\ref{s15})$\sim$(\ref{s16}) with different
Mach number and make comparisons on the following aspects.

\subsubsection{Shock profiles}

In order to facilitate the comparison, let us first introduce
quantities $dP$ and $dQ$ which are defined as the difference of stress and
the heat current between the new version and the conventional N-S
formalism. The definitions are the following: \be \Delta{\mathbb
P}=-\mu\left[{\cal D}_0\ln T+{2\over 3}\nabla\cdot{\bf u}\right]
{\mathbb I}
\ ,
\label{c1}
\ee \be \Delta{{\bf Q}}=-\left(\eta\nabla\ln(\rho T)+{5\over 2}\mu{\cal
D}_0{\bf u}\right) \ .
\label{c2}
\ee For one-D flow, we obtain \be \Delta{\bf P}=\Delta P\hat x\hat x \ ,
\label{c3}
\ee \be \Delta{{\bf Q}}=\Delta Q\hat x\label{c4} \ee with $\Delta P$
and $\Delta Q$ given by \be \Delta P=-\mu\left({\cal D}_0\ln T+{2\over
3}v'\right) \ ,
\label{c5}
\ee \be \Delta Q=-\left\{\eta[\ln(\rho T)]'+{5\over 2}{\cal
D}_0v\right\} \ .
\label{c6}
\ee

In the study of shock wave, the upstream quantities are used to rewrite
the aforementioned dimensionless $dP$ and $dQ$: \be dP\equiv \frac{\Delta
P}{p_1}=-\sqrt{\frac{\gamma}{2}} s M_1\left(
\frac{\theta'}{\theta}+{2\over 3}\omega'\right)\label{c7} \ee \be
dQ\equiv \frac{2\Delta Q}{3p_1v_1}=-{2\over 3}\sqrt{\frac{\gamma}{2}}
 s M_1\left[\frac{15\theta}{4\gamma M_1^2}(\frac{\omega}
{\omega'}-\frac{\theta'}{\theta}+{5\over
2}\omega\omega'\right]\label{c8} \ . \ee From the above definitions, we
see that $dP$ is equivalent to the bulk viscous pressure and $dQ$
accounts for the extra heat current due to the non-adiabatic effect,
both of which are taken as zero in the derivation of N-S
formalism. Such an approximation is not justified when Knudsen number
is not small, as will be shown in the following.
\begin{figure}[hbt]
\begin{center}
\includegraphics[width=6cm,angle=270]{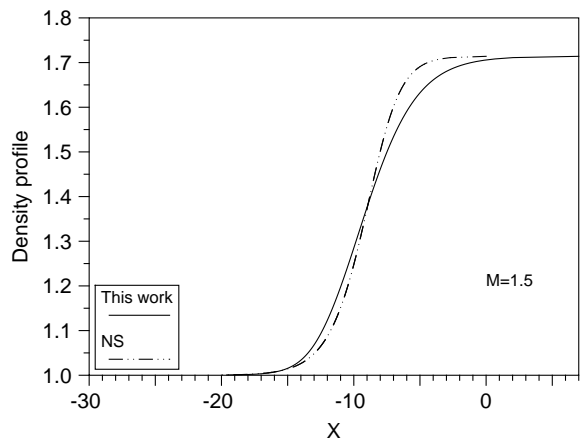}
\end{center}
\caption{Density profiles with Mach number $M$=1.5} \label{fig1}
\end{figure}
\begin{figure}[hbt]
\begin{center}
\includegraphics[width=6cm,angle=270]{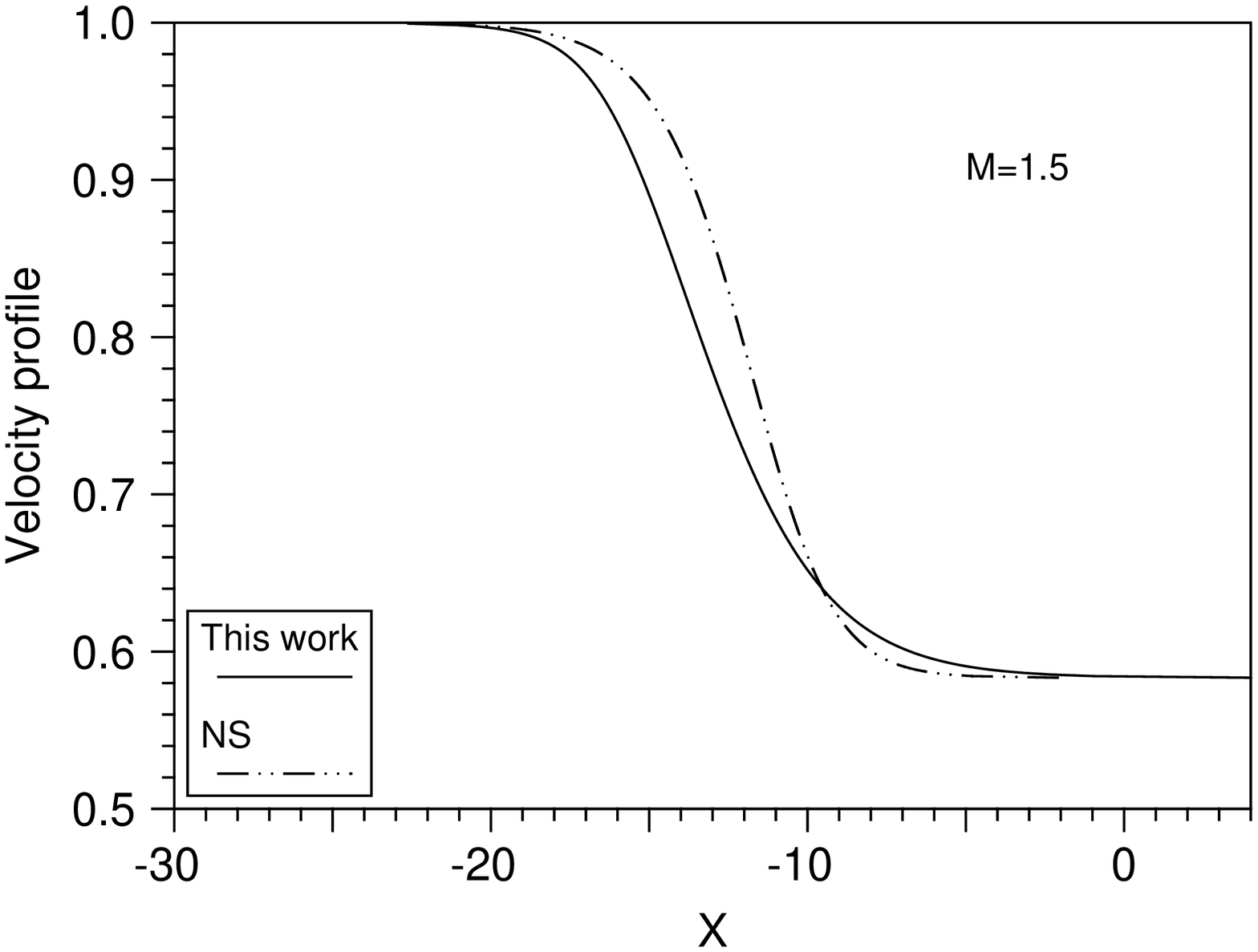}
\end{center}
\caption{Velocity profiles with Mach number $M$=1.5} \label{fig2}
\end{figure}
\begin{figure}[hbt]
\begin{center}
\includegraphics[width=6cm,angle=270]{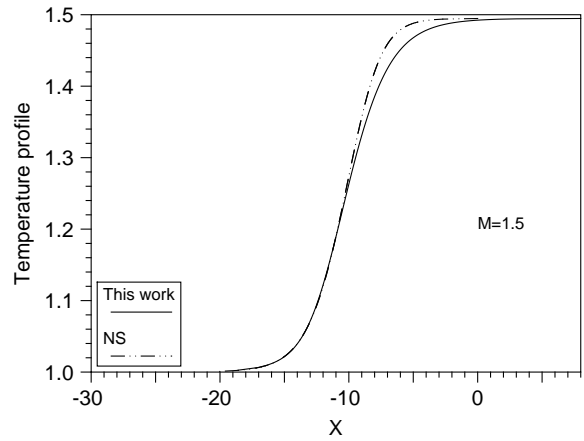}
\end{center}
\caption{Temperature profiles with Mach number $M$=1.5}
\label{fig3}
\end{figure}

Figures~\ref{fig1}$\sim$\ref{fig3} show the density, velocity and
temperature profiles with $M_1=1.5$.  It is seen that the profiles
with the modified hydrodynamics are wider than those with
N-S formalism. Such a trend is a well-known fact and has been verified
experimentally.
\begin{figure}[hbt]
\begin{center}
\includegraphics[width=6cm,angle=270]{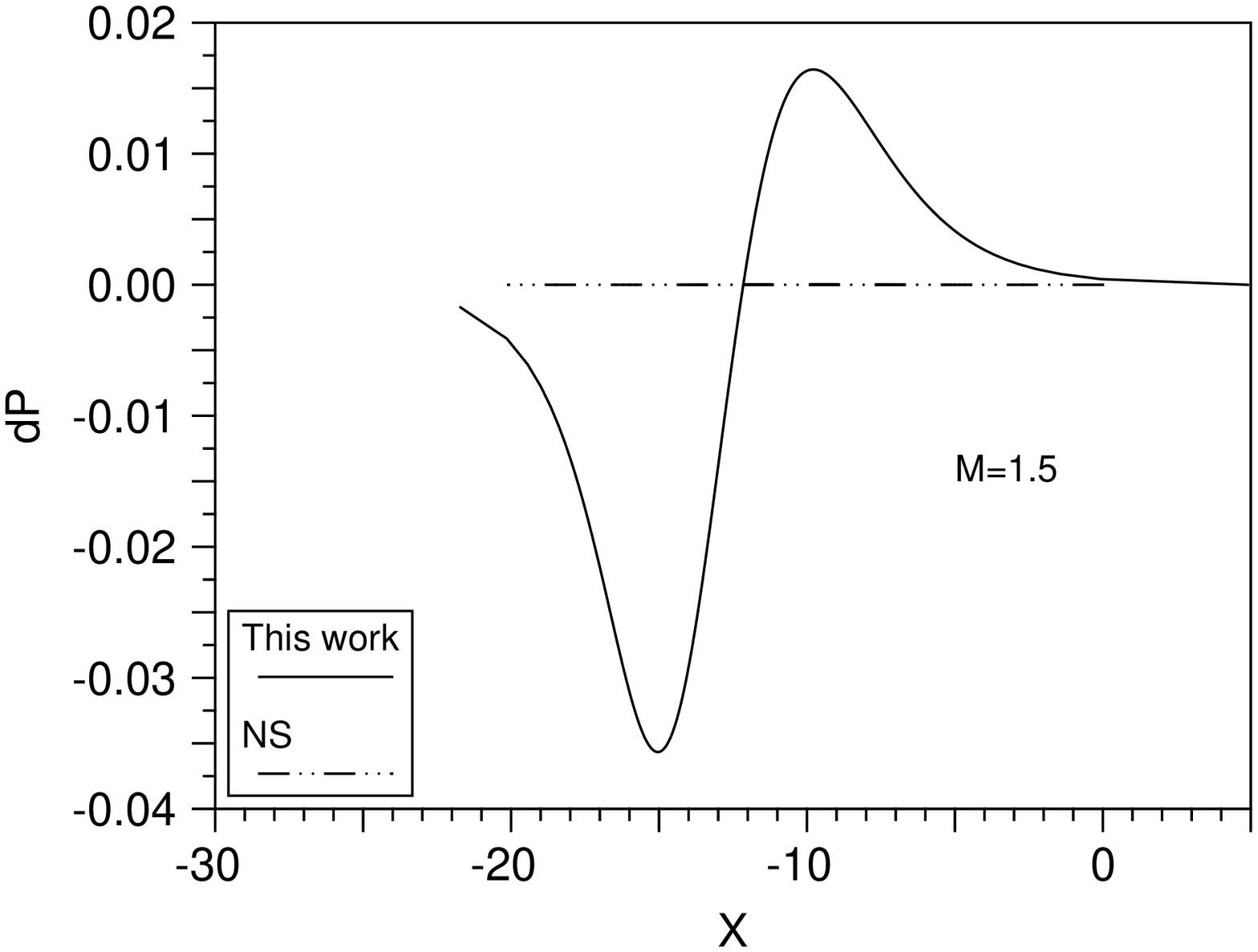}
\end{center}
\caption{$dP$ profiles with Mach number $M$=1.5} \label{fig4}
\end{figure}
\begin{figure}[hbt]
\begin{center}
\includegraphics[width=6cm,angle=270]{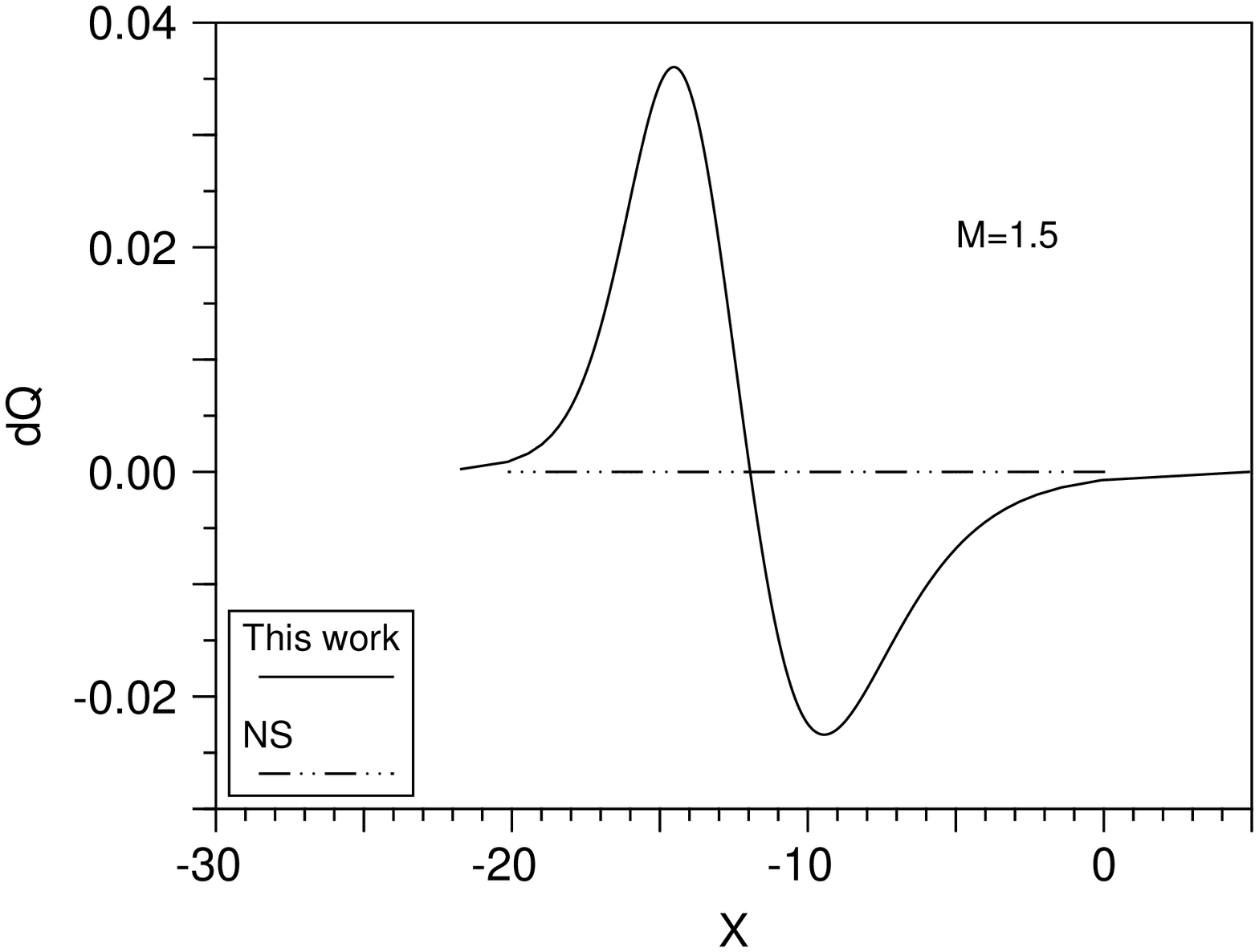}
\end{center}
\caption{$dQ$ profiles with Mach number $M$=1.5} \label{fig5}
\end{figure}

\begin{figure}[hbt]
\begin{center}
\includegraphics[width=6cm,angle=270]{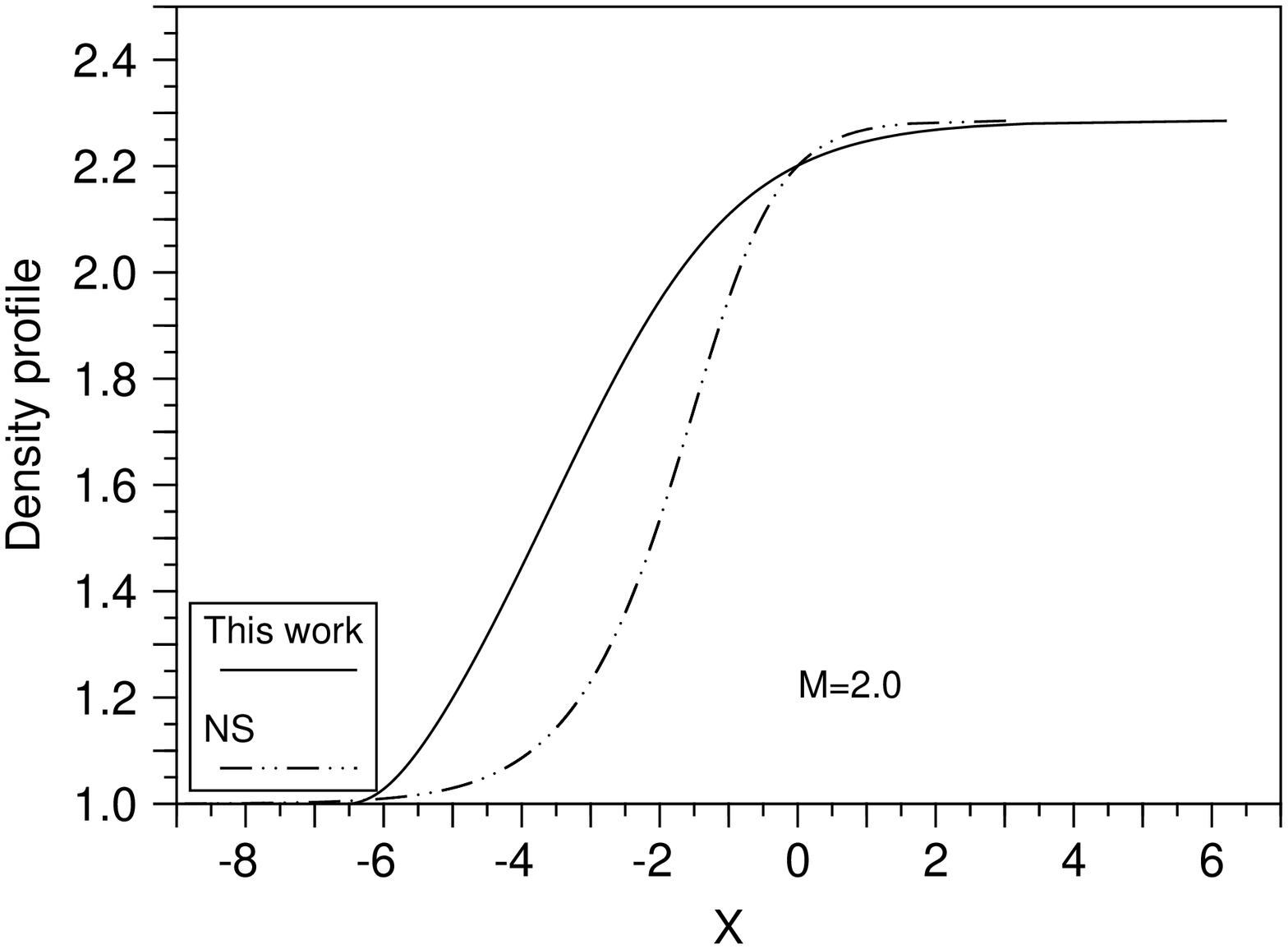}
\end{center}
\caption{Density profiles with Mach number $M$=2.0} \label{fig6}
\end{figure}

The difference in the profiles are due to the nonzero $dP$ and $dQ$,
plotted in Fig.~\ref{fig4} and Fig.~\ref{fig5}, respectively. A simple
estimation shows that the dissipation due to these two extra effects
increases monotonically from the downstream to the upstream, which
tends to lower the gradients of the thermodynamical variables and
results in a wider shock. Such effects become more noticeable when Mach
number is larger, as is shown in Figs.~\ref{fig6}$\sim$\ref{fig10},
where the same quantities as those in Figs.~\ref{fig1}$\sim$\ref{fig5} are
plotted but with $M_1=2$. Compared to the case with $M_1=1.5$, the
profiles are steeper and more asymmetric. The difference of the two
versions of hydrodynamics is now more significant due to the increase
in the Knudsen number with the Mach number. From
Figs.~\ref{fig9}$\sim$~\ref{fig10}, it is clear that the bulk viscous
pressure and the non-adiabatic heat current are no longer negligible.

\begin{figure}[hbt]
\begin{center}
\includegraphics[width=6cm,angle=270]{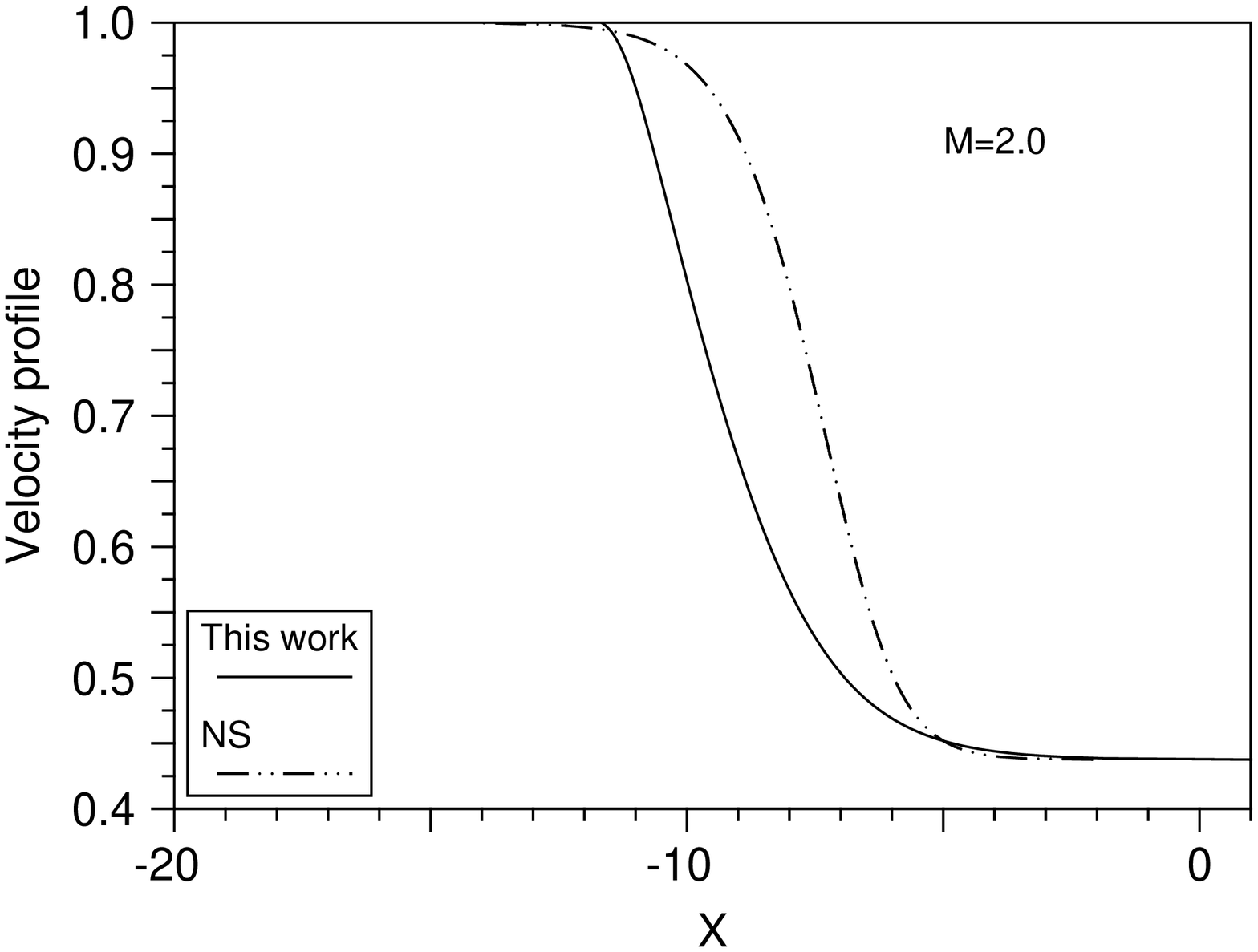}
\end{center}
\caption{Velocity profiles with Mach number $M$=2.0} \label{fig7}
\end{figure}
\begin{figure}[hbt]
\begin{center}
\includegraphics[width=6cm,angle=270]{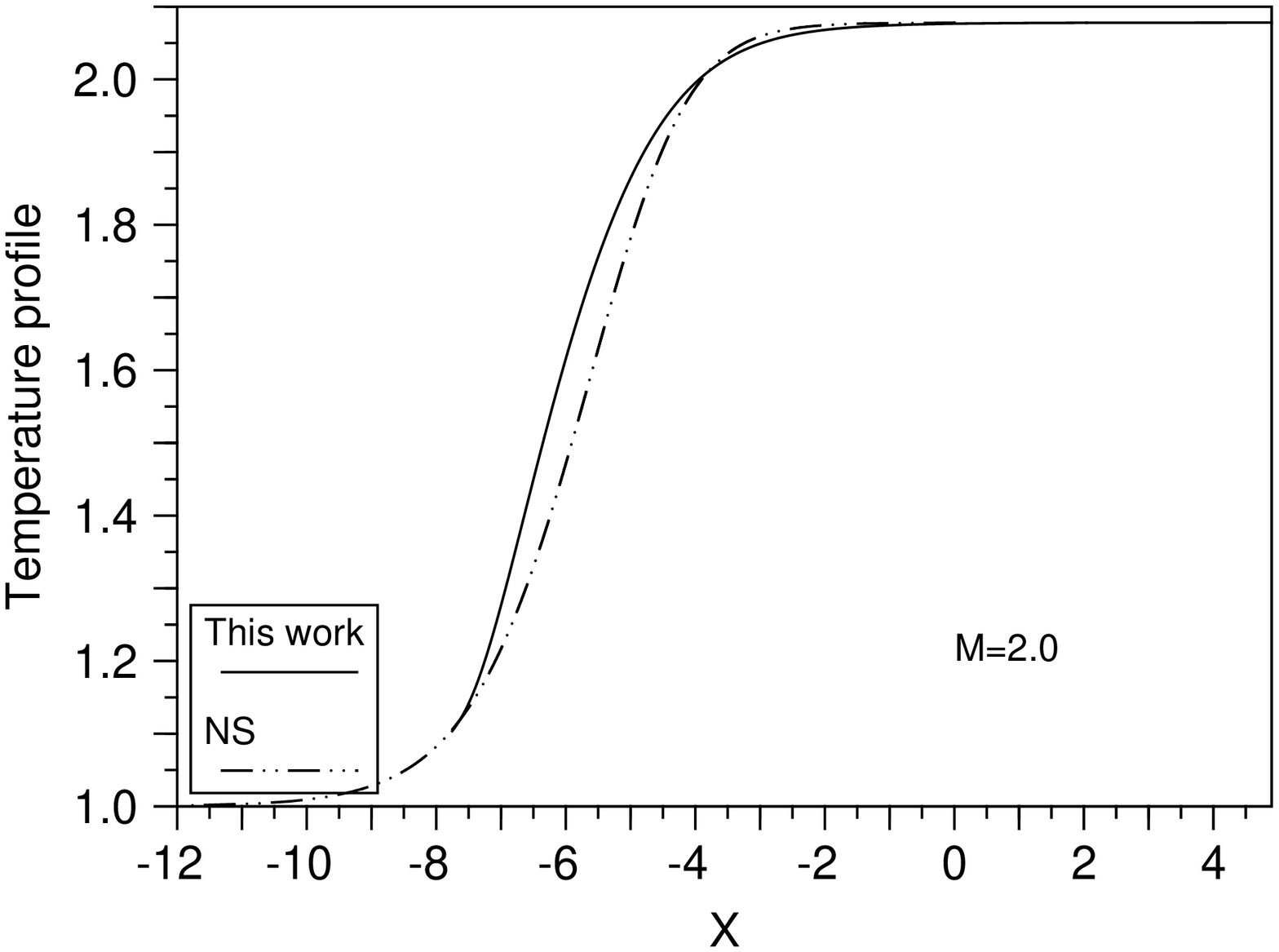}
\end{center}
\caption{Temperature profiles with Mach number $M$=2.0}
\label{fig8}
\end{figure}
\begin{figure}[hbt]
\begin{center}
\includegraphics[width=6cm,angle=270]{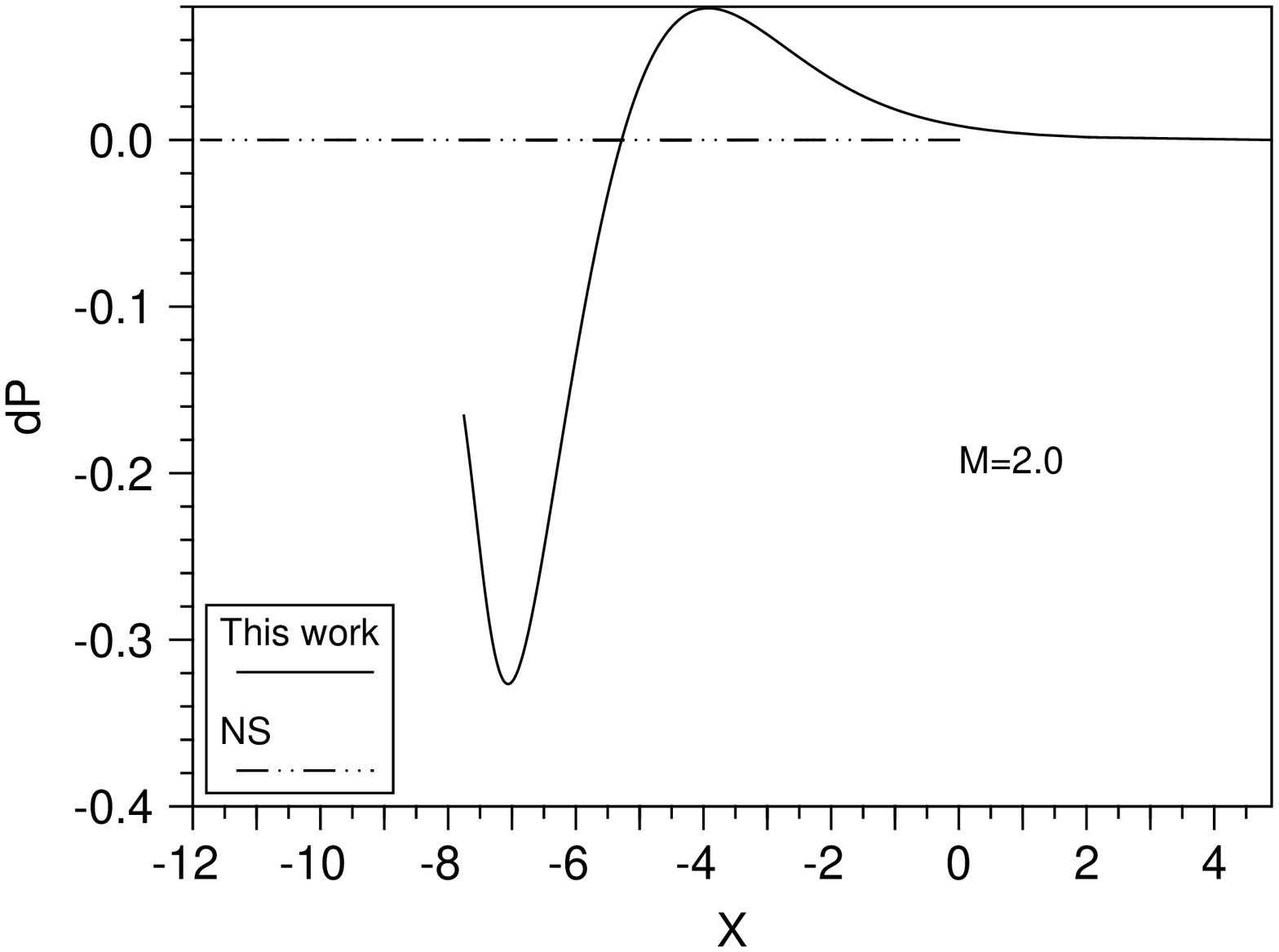}
\end{center}
\caption{$dP$ profiles with Mach number $M$=2.0} \label{fig9}
\end{figure}
\begin{figure}[hbt]
\begin{center}
\includegraphics[width=6cm,angle=270]{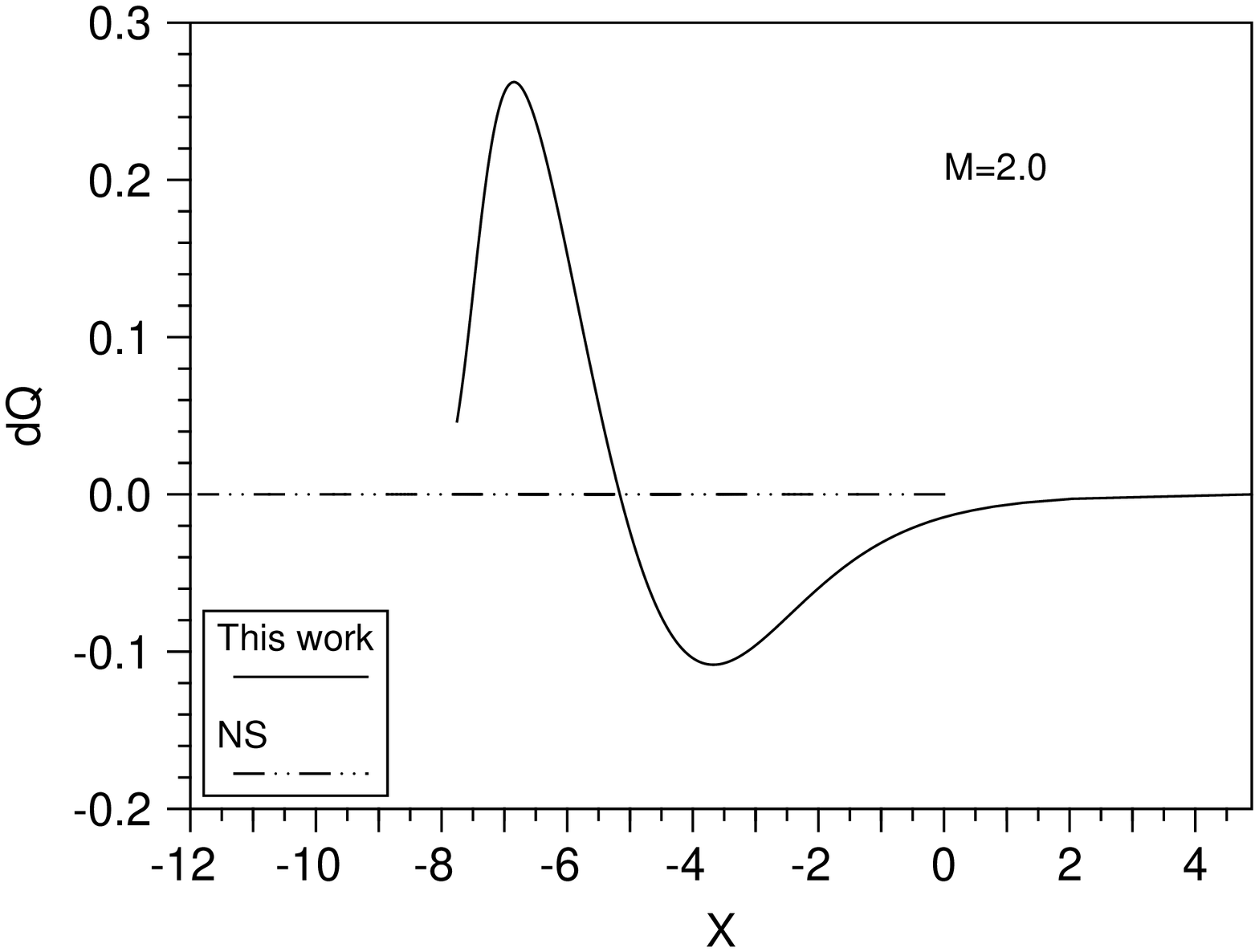}
\end{center}
\caption{$dQ$ profiles with Mach number $M$=2.0} \label{fig10}
\end{figure}

\subsubsection{Inverse thickness, asymmetry and $T$-$\rho$ separation}
 The inverse thickness is defined as \be
 \delta^{-1}=\frac{\rho'_m}{\rho_2-\rho_1} \ ,
\label{thickness}
\ee where $\rho'_m$ is the maximum value of the gradient of density
profile, and as before $\rho_1$ and $\rho_2$ are the upstream and
downstream values, respectively.

 The asymmetry parameter~\cite{fis89} is defined as \be
 A_s=\frac{\int^{\rho_2}_{\rho_a}\rho dx}{\int^{\rho_a}_{\rho_1}\rho
 dx}
\label{As}
\ee with $\rho_a=\frac{1}{2}(\rho_2-\rho_1)$.

Finally, the
temperature-density separation $\Delta_{\rho T}$ is defined as the
distance between the medium value points of temperature and density.

\begin{figure}[hbt]
\begin{center}
\includegraphics[width=6cm,angle=270]{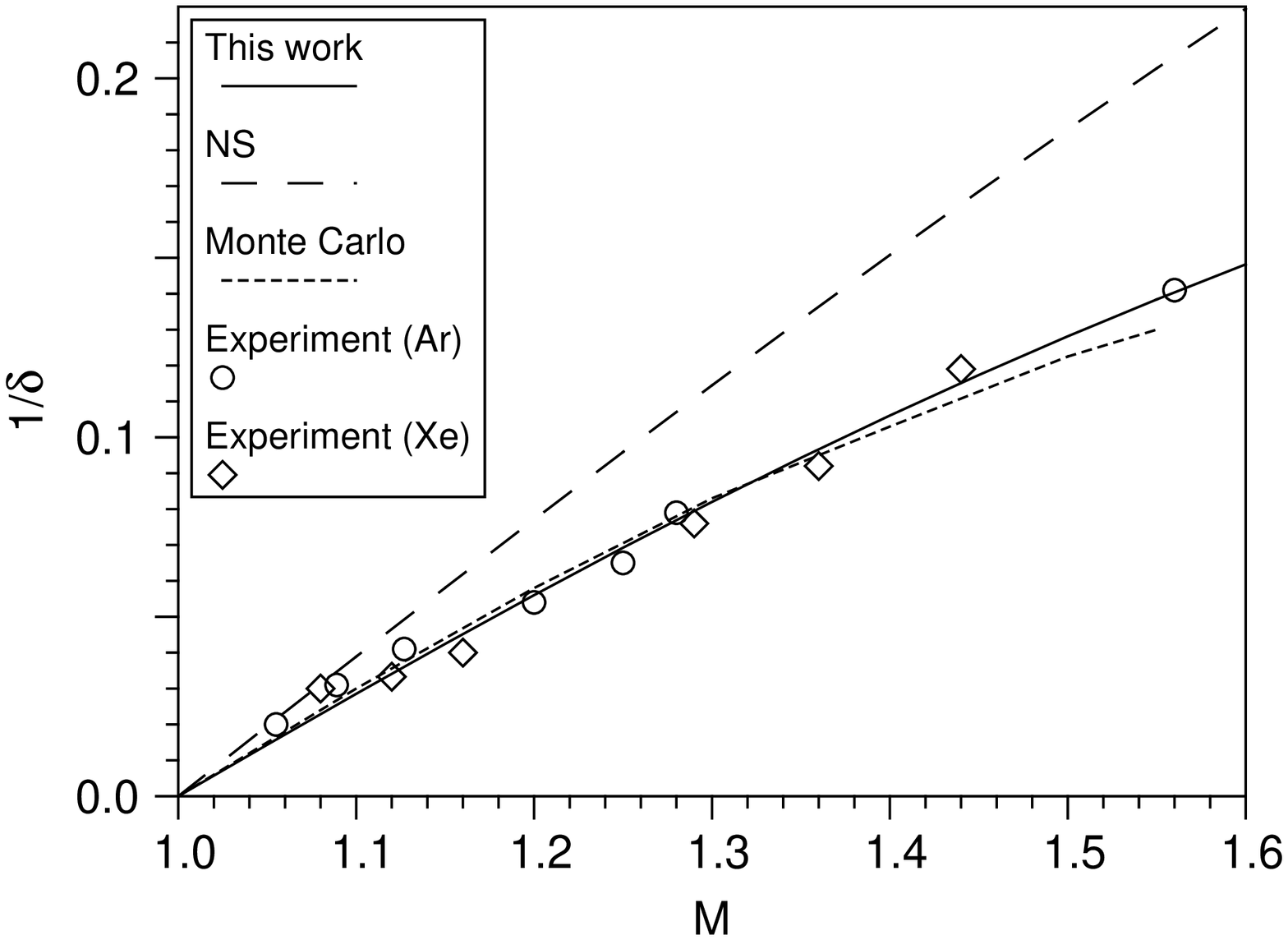}
\end{center}
\caption{Inverse density thickness} \label{fig11}
\end{figure}

Figure~\ref{fig11} shows the inverse thickness versus Mach number. The
experimental data for Argon and Xeron are also shown. Concluded from the plot, the
modified hydrodynamics results in a wider shock for a given Mach
number. In comparison with the experimental data and the direct Monte
Carlo simulation results(for full Boltzmann equation), the accuracy is
significantly improved. The physical explanation for the wider shock is as follows.
The process-dependent bulk viscosity causes extra dissipation, which,
combined with the
non-adiabatic effect, makes the gradients of thermodynamic variables
smoother and thus widens the shock.

As for the asymmetry parameter, in the case of $M_1=2.0$, N-S formalism
gives $A_s=1.25$, the modified one gives $A_s=0.83$, which is closer to
the experimental data $0.93$. Aside from the quantitative improvement, the
modified version also becomes qualitatively consistent with the
experiment, i.e. $A_s$ should be smaller than unity.

For the same Mach number $M_1=2.0$, N-S
leads to a temperature-density separation of $1.04$, whereas the modified
version leads to $1.67$ and the experimental data for Argon is
$1.50$~\cite{fis89}.

\section{Conclusions and Discussions} \label{sec:conclusion}

We have showed that the modified Hilbert expansion leads to a hydrodynamics of
 higher accuracy than that from the classic Chapman-Enskog
expansion. In the study of rapid variation such as high-frequency sound wave propagation, this modified hydrodynamics
provides a result consistent with experiment. The fitting
is uniformly good for all range of Knudsen number. For the study of
shock wave structures, within the range of its critical Mach number,
it provides results comparable to those from direct Monte Carlo
simulation. The physics underlying the new expansion scheme is that it
relaxes the requirement that the gradient terms be small as is the
case in CEE.  This makes the method capable of deriving macroscopic
equations valid for processes far from equilibrium. As we have seen in
the shock structure study, the extra dissipation caused by the new
mechanism of gradient of density smoothes the density profile.  Such a
non-adiabatic dissipation is omitted in the Navier-Stokes formalism
due to the extra constraints enforced by the CEE procedure
in the derivation.  As is discussed in section III, the existence of
the critical Mach
number is caused by the relaxation model we adopted for the kinetic
equation. In one of our following papers~\cite{che00d} we will show that,
when the Boltzmann collisional
integral is used in place of the relaxation term, such a constraint on
Mach number is removed. However, a consistent hydrodynamics
satisfying causality principle yet without extra constraint beyond
relativity on the possible Mach number can only be formulated with the
relativistic Boltzmann equation, as will be shown in Ref.~\cite{che00e}.

\bibliographystyle{prsty}

\end{document}